\documentclass[twocolumn]{aastex61}

\received{}
\revised{}
\accepted{April 11, 2018}
\submitjournal{}

\shorttitle{MHD Simulations of ``Bullet''}
\shortauthors{Nomura et al.}

\begin{document}


\title{Magnetohydrodynamic Simulations of a Plunging Black Hole into a Molecular Cloud}

\correspondingauthor{Mariko Nomura}
\email{mariko.nomura@astr.tohoku.ac.jp}

\author[0000-0002-0786-7307]{Mariko Nomura}
\affil{Department of Physics, Faculty of Science and Technology, 
Keio University, 3-14-1 
Hiyoshi, Kohoku-ku, Yokohama, Kanagawa 223-8522, Japan}
\affil{Astronomical Institute, Tohoku University, Aoba, Sendai 980-8578, Japan}

\author{Tomoharu Oka}
\affiliation{Department of Physics, Faculty of Science and Technology, 
Keio University, 3-14-1 
Hiyoshi, Kohoku-ku, Yokohama, Kanagawa 223-8522, Japan}
\affiliation{School of Fundamental Science and Technology, 
Graduate School of Science and Technology, 
Keio University, 3-14-1 Hiyoshi, Kohoku-ku,
Yokohama, Kanagawa 223-8522, Japan}

\author{Masaya Yamada}
\affiliation{School of Fundamental Science and Technology, 
Graduate School of Science and Technology, 
Keio University, 3-14-1 Hiyoshi, Kohoku-ku,
Yokohama, Kanagawa 223-8522, Japan}

\author{Shunya Takekawa}
\affiliation{School of Fundamental Science and Technology, 
Graduate School of Science and Technology, 
Keio University, 3-14-1 Hiyoshi, Kohoku-ku,
Yokohama, Kanagawa 223-8522, Japan}
\affiliation{Nobeyama Radio Observatory, National Astronomical Observatory of Japan (NAOJ), National Institutes of Natural Sciences (NINS), 462-2 Nobeyama, Minamimaki, Minamisaku-gun, Nagano 384-1305, Japan}

\author{Ken Ohsuga}
\affiliation{National Astronomical Observatory of Japan, 
Osawa, Mitaka, Tokyo 181-8588, Japan}
\affiliation{School of Physical Sciences, 
Graduate University of Advanced Study (SOKENDAI),
Shonan Village, Hayama, Kanagawa 240-0193, Japan}

\author{Hiroyuki R. Takahashi}
\affiliation{National Astronomical Observatory of Japan, 
Osawa, Mitaka, Tokyo 181-8588, Japan}

\author{Yuta Asahina}
\affiliation{National Astronomical Observatory of Japan, 
Osawa, Mitaka, Tokyo 181-8588, Japan}








\begin{abstract}
Using two-dimensional magnetohydrodynamic simulations, 
we investigated the gas dynamics around 
a black hole plunging into a molecular cloud.
In these calculations, we assumed a parallel-magnetic-field layer in the cloud.
The size of the accelerated region is far larger than the Bondi--Hoyle--Lyttleton radius,
being approximately inversely proportional to the Alfv\'en Mach number for the plunging black hole.
Our results successfully reproduce the ``Y'' shape in position--velocity maps of the ``Bullet''
in the W44 molecular cloud.
The size of the Bullet is also reproduced within an order of magnitude 
using a reasonable parameter set.
This consistency supports the shooting model of the Bullet, 
according to which an isolated black hole plunged into a molecular cloud to form a compact 
broad-velocity-width feature.

\end{abstract}



\section{Introduction} \label{sec:intro}
To date, $\sim\! 60$ stellar mass black hole (BH) candidates have been detected in our Galaxy by X-ray observations \citep{2016A&A...587A..61C}, while their total number is estimated to be $\sim \! 10^8$--$10^9$ \citep{2002MNRAS.334..553A, 2017MNRAS.468.4000C}.  
This sharp discrepancy is due to the extremely low percentage of BHs in close binary systems, in which abundant mass accretion from companion stars activate them.  Therefore, almost all BHs in our Galaxy still remain undetected. 

An isolated, inactive BH would pull up ambient material, leaving a trace in the interstellar medium as a spatially compact, broad-velocity-width feature.
The ``Bullet'' in the W44 molecular cloud is a candidate for such a BH trace \citep{2013ApJ...774...10S}.  W44 is a supernova remnant (SNR) interacting with an adjacent giant molecular cloud \citep{1997ApJ...489..143C,1998ApJ...505..286S,2004AJ....127.1098S,2013ApJ...774...10S}. 
In the process of investigating the gas kinematics of the W44 molecular cloud, we noticed an extraordinary broad-velocity-width feature \citep{2013ApJ...774...10S}, the Bullet. 
Follow-up observations by \citet[][hereafter Y17]{Yamada17} have revealed its compact appearance ($0.5\times 0.8\,{\rm pc}^2$), broad-velocity width nature ($\Delta V\!\sim\!100$), and unique ``Y'' shape in the position--velocity maps.

The Bullet is intense in CO {\it J}=4--3 and HCN {\it J}=1--0 emissions, suggesting that it consists of warm and dense molecular gas.  
The total kinetic energy of the Bullet is $\sim\!10^{48}\,{\rm erg}$. 
This is approximately 1.5 orders of magnitude greater than the kinetic energy of a supernova 
sharing the small solid angle of the Bullet with respect to the W44 center.  
Y17 proposed two scenarios of Bullet formation: (1) the expansion model and (2) the shooting model.  
Both scenarios assume an isolated BH that contributes to the formation of the Bullet. 
In the expansion model, 
an additional explosive event triggered by mass accretion onto an isolated BH accounts for 
the kinematics of the Bullet, 
but the conversion process from the gravitational energy to the kinetic energy is unclear. 
The shooting model seems to be more plausible. 
In this model, the plunge of a $\gtrsim \! 30\,M_\odot$ BH into the high-density layer toward us 
successfully explains the broad velocity width as well as the enormous kinetic energy of the Bullet
(see Figure 3(b) in Y17). 
The gas dragged by the plunging BH might correspond to the ``Y'' shape on the position--velocity map.


One problem is that a native shooting model cannot reproduce the spatial size of the Bullet.  This is because the size of the accretion zone, which may correspond to the Bullet size, is described by the Bondi--Hoyle--Lyttleton (BHL) radius \citep{1939PCPS...35..405H,1944MNRAS.104..273B,2004NewAR..48..843E}, $R_{\rm BHL}=2GM_{\rm BH}/(c_{\rm s}^2+v^2)$,
where $G$, $M_{\rm BH}$, $c_{\rm s}$, and $v$ are the gravitational constant, the BH mass, the sound speed, and the velocity of the plunging BH.  When $M_{\rm BH}\sim 30 \,M_\odot$, $c_{\rm s}\sim 1\,{\rm km\,s^{-1}}$, and $v \sim 100\,{\rm km\,s^{-1}}$, the BHL radius becomes $\sim\! 3\times 10^{-5}\,{\rm pc}$, which is too small to reproduce the Bullet size ($\sim\!0.5 {\rm pc}$).

In this paper, we examine the effect of the magnetic field on the size of the accretion zone around a plunging BH. 
The magnetic field is frozen in the partially ionized interstellar gas, 
and the bent shape of the field lines propagates with the Alfv\'en speed.
If the BH plunges into the parallel-magnetic-field layer and the magnetic field lines are caught on the gas 
in the vicinity of the BH, 
the gas outside of $R_{\rm BHL}$ would be dragged by the plunging BH via the magnetic tension force. 
This effect may enlarge the size of the accretion zone 
if the magnetic field is strong enough, 
which is the case in the W44 expanding shell ($B\!\sim\!500\,\mu{\rm G}$; \citealt{Hoffman05}). 
In this case, the lower part of the ``Y'' shape in the position--velocity map may correspond to 
the accelerated gas in front of the BH, which is moving with the plunging speed of BH.
The upper part of the ``Y'' shape can be interpreted as the gas around the BH accelerated 
from the magnetic tension force, 
whose velocity may decrease with the distance from the BH. 

In order to quantitatively investigate the magnetohydrodynamic (MHD) effect in the shooting model, 
we simulated the plunging of a BH into a parallel-magnetic-field layer by considering a large-scale gas flow around a stationary BH. 
A two-dimensional MHD code was employed. 
In Section \ref{sec:methods}, we explain the calculation method. 
The results of the simulations are shown in Section \ref{sec:results}.  
Sections \ref{sec:discussions} and \ref{sec:conclusions} are devoted to discussions and conclusions.

\section{Basic Equations and Models}\label{sec:methods}
We calculate the gas dynamics 
around a plunging BH 
using MHD simulations 
in Cartesian coordinates $(x,\,y,\,z)$.
In our simulations, 
a BH is located at the center of the coordinate system
and ambient gas flows in with high velocity.
The basic equations of ideal MHD are
\begin{equation}
\frac{\partial\rho}{\partial t}+\nabla\cdot(\rho \mbox{\boldmath $v$})=0,
\end{equation}
\begin{equation}
\frac{\partial\rho \mbox{\boldmath $v$}}{\partial t}+\nabla\cdot \left(\rho \mbox{\boldmath $v$} \otimes \mbox{\boldmath $v$} +p+ \frac{B^2}{8\pi}-\frac{\mbox{\boldmath $B$} \otimes \mbox{\boldmath $B$} }{4\pi}\right)=-\rho\frac{GM_{\rm BH}\mbox{\boldmath $r$}}{r^3},
\end{equation}
\begin{equation}
\frac{\partial}{\partial t} \left( e+\frac{B^2}{8\pi} \right)+
\nabla\cdot \left[ \left( e+p \right)\mbox{\boldmath $v$}
-\frac{\left( \mbox{\boldmath $v$} \times \mbox{\boldmath $B$} \right)\times \mbox{\boldmath $B$} }{4\pi}\right]
=0,
\end{equation}
\begin{equation}
\frac{\partial \mbox{\boldmath $B$}}{\partial t}-\nabla\times(\mbox{\boldmath $v$}\times \mbox{\boldmath $B$})=0,
\end{equation}
where, $\rho$, {\boldmath $v$}, $p$, {\boldmath $B$}, {\boldmath $r$}, and $e$
are the mass density, velocity, pressure, magnetic field, distance from the BH, 
and energy density of the gas written by $e=p/(\gamma-1)+\rho v^2/2$ with $\gamma=5/3$.
In our calculations, 
we assume that the energy dissipation in shock has a time scale comparable to that of the cooling.  
At each time step, we reset the pressure and the energy density of the gas using an isothermal equation of state,
$p=\rho c_{\rm s}^2$, 
where we set a sound speed to $c_{\rm s}=0.91\,{\rm km\,s^{-1}}$, 
which is consistent with the observed temperature of the Bullet ($T=100\,{\rm K}$, Y17) 
and a mean molecular weight of 1.0. 
The last term of the equation of motion is 
the gravitational force 
due to the BH located at the origin. 

The computational domain is $-0.5\,{\rm pc} \leq x \leq 0.5\,{\rm pc}$ 
and $-0.5 \,{\rm pc}\leq y \leq 0.5\,{\rm pc}$. 
We prepare $720\times 720$ grids in the domain for $M_{\rm BH} = 10^4\,M_\odot$ so as to set 7--8 grids 
in the BHL radius, $R_{\rm BHL}$.
The resolution is $\sim\! 1.4\times 10^{-3}\, {\rm pc}$.
We also employ $M_{\rm BH} = 10^3 \,M_\odot$, $10^{3.5}\,M_\odot$, and $10^{4.5}\,M_\odot$. 
For these cases, we divide the domain into $7200\times 7200$, $2160\times 2160$, and $240\times 240$ grids,
corresponding to resolutions of $\sim\! 1.4\times 10^{-4}\, {\rm pc}$, $\sim\! 4.6\times 10^{-4}\,{\rm pc}$, and $\sim\! 4.2\times 10^{-3}\,{\rm pc}$.
The choices of $M_{\rm BH}$ larger than $10^3$ $M_{\odot}$ are for resolving the BHL radii with the realistic number of computational grids.
After performing simulations with four $M_{\rm BH}$, we check the $M_{\rm BH}$-dependence on the size of the accelerated region 
(Section \ref{subsec:s}, Figure \ref{fig6}) 

%

\begin{figure}[ht!]
\plotone{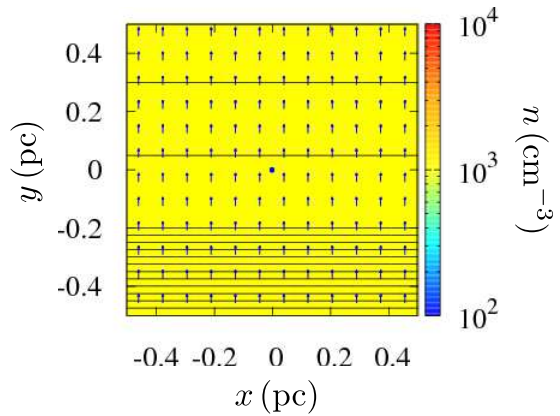}
\caption{Initial conditions for $M_{\rm BH}=10^4 \,M_\odot$, 
$v_{y0}=50\,{\rm km\,s^{-1}}$, 
$n_0=10^3\,{\rm cm^{-3}}$, 
and $B_{x0}=500\,\mu {\rm G}$.
Color map shows the density in the $x$-$y$ plane. 
Black lines and blue arrows show the magnetic field lines and the velocity field. 
The BH is located at the center of the coordinate system.
}
\label{fig0}
\end{figure}

We show the initial condition of the fiducial model in Figure \ref{fig0}.
In the region $r \geq R_{\rm BHL}$, 
the number density and the velocity are set to $n=n_0$ and $\mbox{\boldmath $v$}=(0,v_{y0},0)$.
We suppose the situation that a BH plunges into a magnetized layer (high-$B$ layer)
with a relative velocity of $\sim\!100\,{\rm km\,s^{-1}}$.
We reproduce this situation by shedding high-$B$ layer to a stationary BH, 
because it is difficult to treat a BH moving in the computational box.  
Since the high-$B$ layer is accelerated by the magnetic pressure before reaching the BH 
(see also Section \ref{sec:fiducial}),
we choose the initial velocity ($v_{y0}$) so that the plunge velocity becomes $100\,{\rm km\,s^{-1}}$.
For the fiducial model, the initial velocity is set to $v_{y0}=50\,{\rm km\,s^{-1}}$. 
We employ $n_0=10^3\,{\rm cm^{-3}}$ to simulate the high-$B$ (density) layer after the supernova blast wave.  
This value is between the typical density in molecular clouds and that in the Bullet (Y17).
We also perform the simulations for different $n_0$ and $v_{y0}$ 
(see Section \ref{subsec:p} and Section \ref{subsec:s}).

In the region $r < R_{\rm BHL}$, 
the density is set to $n=0.1n_0$ and the velocity to zero.
On the upstream side ($y\leq -0.2\,{\rm pc}$), 
the magnetic field is set to $\mbox{\boldmath $B$}=(B_{x0},0,0)$, 
which describes 
the high-$B$ layer.
We employ $B_{x0}=500\,\mu {\rm G}$ based on the previous measurement in the W44 expanding shell \citep{Hoffman05}.
Sections \ref{subsec:p} and \ref{subsec:s} describe the other cases.
In the downstream side, the magnetic field is set to $\mbox{\boldmath $B$}=(50\,\mu{\rm G},0,0)$.
 
We apply the free boundaries at $x=-0.5\,{\rm pc}$, $x=0.5\,{\rm pc}$, and $y=0.5\,{\rm pc}$. 
At the boundary $y=-0.5\,{\rm pc}$, the density, $x$-component of the magnetic field, and $y$-component of the velocity are kept constant $n=n_0$, $B_x=B_{x0}$, and $v_y=v_{y0}$. 
We impose the free boundary conditions for the other variables.
This means that the gas is initially flowing, maintaining the initial density and magnetic field strength in $y\leq -0.5\,{\rm pc}$.  
This situation corresponds to a high-velocity plunging of a BH into a uniform gas cloud.  
In $r<R_{\rm BHL}$, the velocity is fixed to zero so as to reproduce the condition that the gas is trapped by the BH and it traps the magnetic field lines.
The density is also kept constant at $n=0.1n_0$ so that the gas flowing into this region accretes onto the BH and does not appear 
in the computational domain again.

Numerical simulations are carried out by using the MHD code CANS+ \citep{Matsumoto16}.
 The code employs the HLLD approximate Riemann solver \citep{Miyoshi05}. 
We apply second-order spatial accuracy 
by using monotone upstream-centered schemes for conservation laws \citep[MUSCL,][]{vanLeer79}. 
A third order TVD Runge--Kutta scheme is used for solving the time integration. 
A hyperbolic divergence cleaning method is employed \citep{Dedner02}.


\section{Results}\label{sec:results}
\subsection{Fiducial Model}
\label{sec:fiducial}
\begin{figure}[ht!]
\plotone{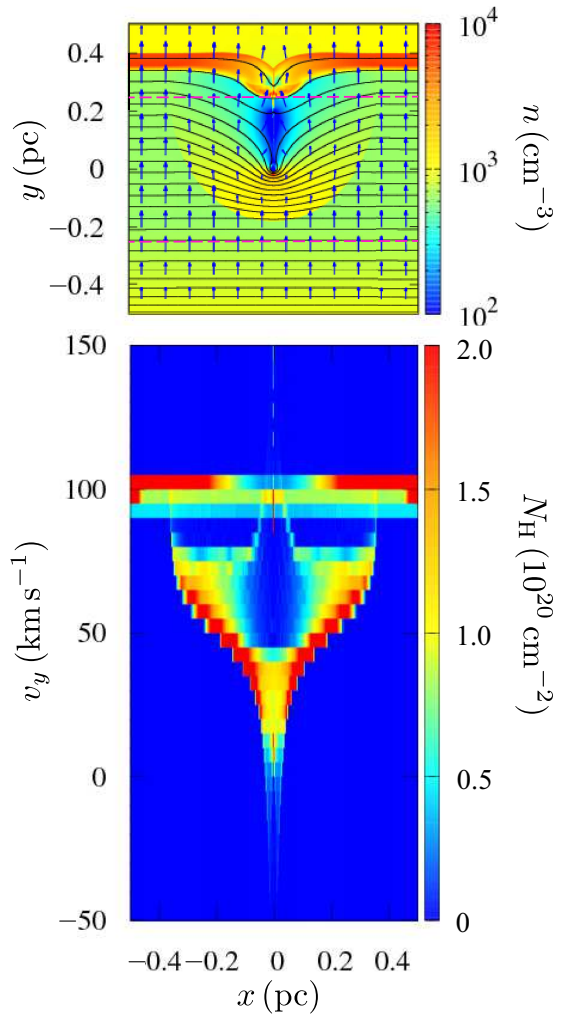}
\caption{Density map (top panel) in the $x$-$y$ plane 
and column density map in the $x$-$v_y$ plane (bottom panel) 
at $t=5.38\times 10^3 \,{\rm yr}$ for $M_{\rm BH}=10^4 \,M_\odot$, 
$v_{y0}=50\,{\rm km\,s^{-1}}$, 
$n_0=10^3\,{\rm cm^{-3}}$, 
and $B_{x0}=500\,\mu {\rm G}$.
In the top panel, 
black lines and blue arrows show the magnetic field lines and the velocity field. 
Magenta dashed lines show the range of integration. 
The BH is located at the center of the coordinate system.
}
\label{fig1}
\end{figure}

Figure \ref{fig1} shows the results for $M_{\rm BH}=10^4 \,M_\odot$, 
$v_{y0}=50\,{\rm km\,s^{-1}}$, 
$n_0=10^3\,{\rm cm^{-1}}$, 
and $B_{x0}=500\,\mu {\rm G}$ 
(the fiducial model).
In this case, the gas is accelerated to $v_y\sim 100\,{\rm km\,s^{-1}}$ by the magnetic pressure before the 
high-$B$ layer reaches the BH. 
This velocity corresponds to the plunging speed of the BH in the shooting model
(hereafter we call it the inflow velocity, $v_{\rm in}$).
The top panel shows the density map at $t=5.38\times 10^3 \,{\rm yr}$ 
when the high-$B$ layer passes $\sim\! 0.4\,{\rm pc}$ 
after reaching the BH. 
%

In the region $|x|\gtrsim 0.36\,{\rm pc}$, the gas flows as it is not affected by the magnetic tension force.
%
Near the BH ($|x|\lesssim 0.36\,{\rm pc}$), 
the high-density and low-velocity region 
(accelerated region) 
appears and forms an arcuate shape at the side facing the BH
(see the yellow region at $-0.2\,{\rm pc}\lesssim y \lesssim0.2\,{\rm pc}$). 
At the front of the BH,
gas is compressed 
and the velocity is decelerated 
(accelerated toward the negative direction).
In this area, the magnetic field is enhanced to $\sim\! 600$--$3000\,\mu {\rm G}$.
This is because 
gas is frozen into the magnetic field and 
the field lines are 
caught on the gas dammed up in $r <R_{\rm BHL}$.
At the front of the BH, the magnetic pressure strongly accelerates the gas, 
and at both sides of the BH, 
the magnetic tension force due to the curved magnetic field lines contributes to the acceleration.
The following flow collides with the accelerated flow inducing shock.

The bottom panel of Figure \ref{fig1}
shows a position--velocity ($x$-$v_y$) map. 
We divide the $v_y$-axis into grids of the width $\Delta v_y =5\,{\rm km\,s^{-1}}$. 
The color represents the column density of the gas in each velocity grid. 
The column density is integrated along the $y$-axis over the range $-0.25\,{\rm pc}\leq y \leq 0.25 \,{\rm pc}$
(between the magenta dashed lines in the top panel of Figure \ref{fig1})
in order to focus on the velocity structure around the BH.

In the region far from the BH ($|x|\gtrsim 0.36\,{\rm pc}$),
the bulk of the gas has a velocity of $\sim\! 100\,{\rm km\,s^{-1}}$, 
because of less influence of 
the magnetic field 
at this time.
In the region $|x|\lesssim 0.36\,{\rm pc}$, 
another velocity component appears at $ v_y\lesssim 100\,{\rm km\,s^{-1}}$.
The $ v_y\sim 100\,{\rm km\,s^{-1}}$ component is contributed by the less dense gas 
in the upstream side ($y\lesssim -0.18\,{\rm pc}$). 
%
The low-velocity component ($v_y<100\,{\rm km\,s^{-1}}$) makes a ``Y'' shape on the $x$-$v_y$ map.
The widths of the ``Y'' shape in the $x$-direction ($d$) and in the $v_y$-direction ($\Delta v_y$) are 
$\sim\! 0.72\,{\rm pc}$ and $\sim\! 115\,{\rm km\,s^{-1}}$ respectively.
This component is contributed by the dense gas in the accelerated region around the BH. 
%
The $v_y<0$ component is contributed by the gas in the vicinity of the BHL radius,
which falls into the BH along the magnetic field. 


\begin{figure}[ht!]
\plotone{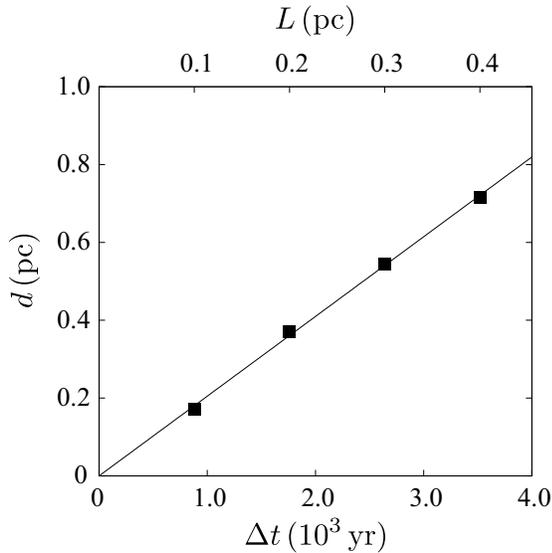}
\caption{
Time evolution of the size of the ``Y'' shape on the $x$-$v_y$ map.
Black squares show the results of the fiducial model
when $\Delta t=8.80\times 10^2$, $1.76\times 10^3$, $2.64\times 10^3$, and $3.52\times 10^3 \,{\rm yr}$ 
($L=0.1$, $0.2$, $0.3$, and $0.4\,{\rm pc}$).
The best-fitting line, $d/{\rm pc}=2.05\times10^{-4}\Delta t/{\rm yr}$ ($d=1.8L$), 
is shown by the solid line.
}
\label{fig2}
\end{figure}

Figure \ref{fig2} shows the time evolution of the size of the accelerated region,
which corresponds to the size of the ``Y'' shape on the $x$-$v_y$ map, $d$.
Here, $\Delta t$ is the time measured from the moment 
that the high-$B$ layer reached the BH.
Using the speed of the flow ($\sim\! 100\,{\rm km\,s^{-1}}$), 
the time is converted to the width of the layer that passed through the BH, $L$.
The size increases in proportion to $\Delta t$ and $L$.
The best-fitting line (the solid black line) is $d/{\rm pc}=2.05\times10^{-4}\Delta t/{\rm yr}$ ($d=1.8L$).
This shows that the accelerated region expands with the Alfv\'en speed at the vicinity of $R_{\rm BHL}$, 
$\sim\! 200\,{\rm km\,s^{-1}}$, which is roughly $5.8$ times the Alfv\'en speed in the high-B layer.

\subsection{Parameter Dependence}
\label{subsec:p}
\begin{figure*}[ht!]
\plotone{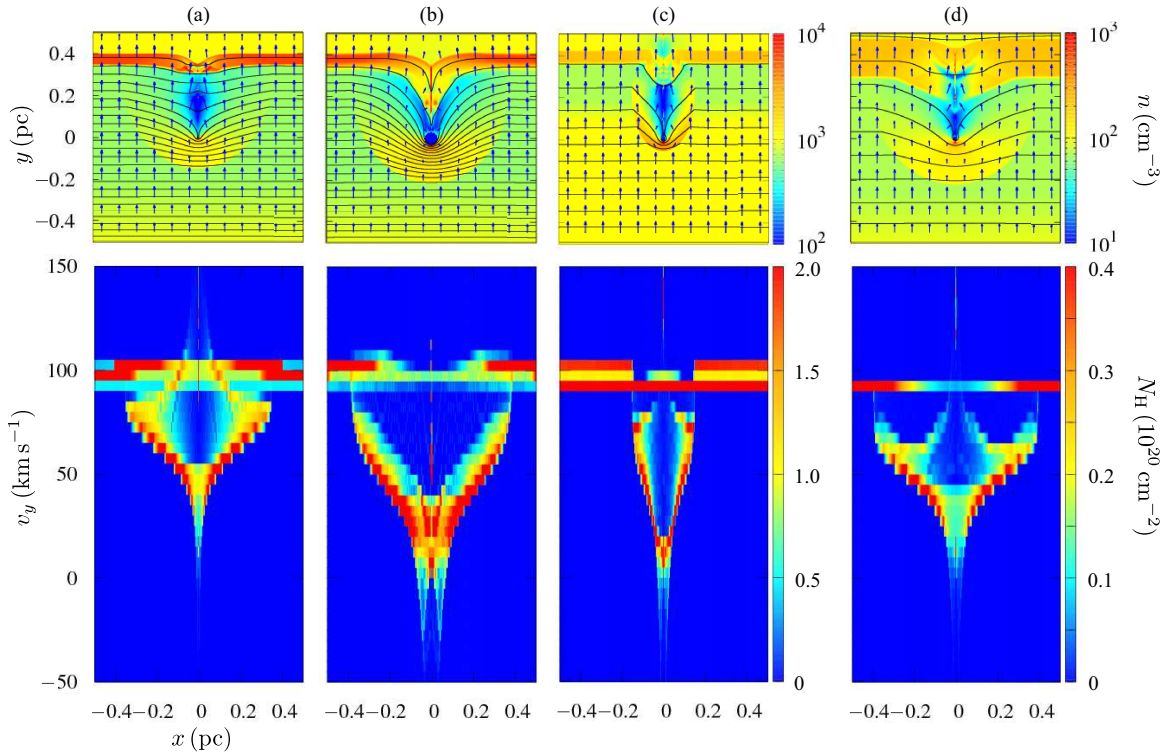}
\caption{
Same as Figure \ref{fig1} but for 
(a) $M_{\rm BH}=10^{3.5}M_\sun$, (b) $M_{\rm BH}=10^{4.5}M_\sun$, (c) $B_{x0}=158\,\mu{\rm G}$, and (d) $B_{x0}=158\,\mu{\rm G}$ and $n_0=100\,{\rm cm^{-3}}$.}
\label{fig3}
\end{figure*}

Each panel of Figure \ref{fig3} is the same as that of Figure 1, 
but the parameters are different from the fiducial model.
Figures \ref{fig3}(a) and \ref{fig3}(b) show the results in the case of
$M_{\rm BH}=10^{3.5}\,M_\odot$ and $10^{4.5}\,M_\odot$.
In both figures, the overall structures are quite similar to the result of the fiducial model.  
In Figure \ref{fig3}(a), 
the size and the velocity width of the ``Y'' shape in the $x$-$v_y$ map
are $d \sim 0.70\,{\rm pc}$ and $\Delta v_y \sim 100\,{\rm km\,s^{-1}}$ respectively. 
In Figure \ref{fig3}(b), these values are $d \sim 0.78\,{\rm pc}$ and $\Delta v_y \sim 130\,{\rm km\,s^{-1}}$.
These results show that 
the position--velocity structure is almost independent of the BH mass (see also Figure \ref{fig1}).

Figure \ref{fig3}(c) shows the results for the weak magnetic field, $B_{x0}=158\,\mu{\rm G}$.
In order to adjust the inflow velocity to $v_{\rm in}\sim 100\,{\rm km\,s^{-1}}$, 
which is the same as the other cases,
we set the initial velocity to $v_{y0}=90\,{\rm km \,s^{-1}}$. 
The other parameters are the same as those of the fiducial model.
The density map shows that the accelerated region around the BH is small, $|x|\lesssim 0.15\,{\rm pc}$. 
For this parameter set, the magnetic field lines are sharply bent by the ram pressure of the flow, since the magnetic field strength is weak.
As a result, in the $x$-$v_y$ map, the size of the ``Y'' shape is $d \sim 0.3\,{\rm pc}$.
The velocity width of the ``Y'' shape is almost the same as that of the fiducial model, 
but the matter located at $x\sim 0$ is concentrated in the narrow velocity range 
$-15\,{\rm km\,s^{-1}}$ to $+ 15\,{\rm km\,s^{-1}}$.
This is because the gradient of the magnetic field strength is large at the front of the BH and the flow is rapidly decelerated.

Figure \ref{fig3}(d) is the same as Figure \ref{fig1} except that $B_{x0}=158\,\mu{\rm G}$ and $n_0=100\,{\rm cm^{-3}}$. 
The size of the accelerated region is larger than that of Figure \ref{fig3}(c) and comparable to that of Figure \ref{fig1}. 
In this case, the magnetic field is weak, but the ram pressure of the flow is also small due to the low density. 
As a consequence, the magnetic field lines draw gentle curves near the BH.
In the $x$-$v_y$ map, the column density in each cell is smaller than that of the fiducial model, 
but the position--velocity structure is very similar to that of Figure \ref{fig1}.

\section{Discussions}\label{sec:discussions}
\subsection{Size of Accelerated Gas}\label{subsec:s}
\begin{figure*}[ht!]
\plotone{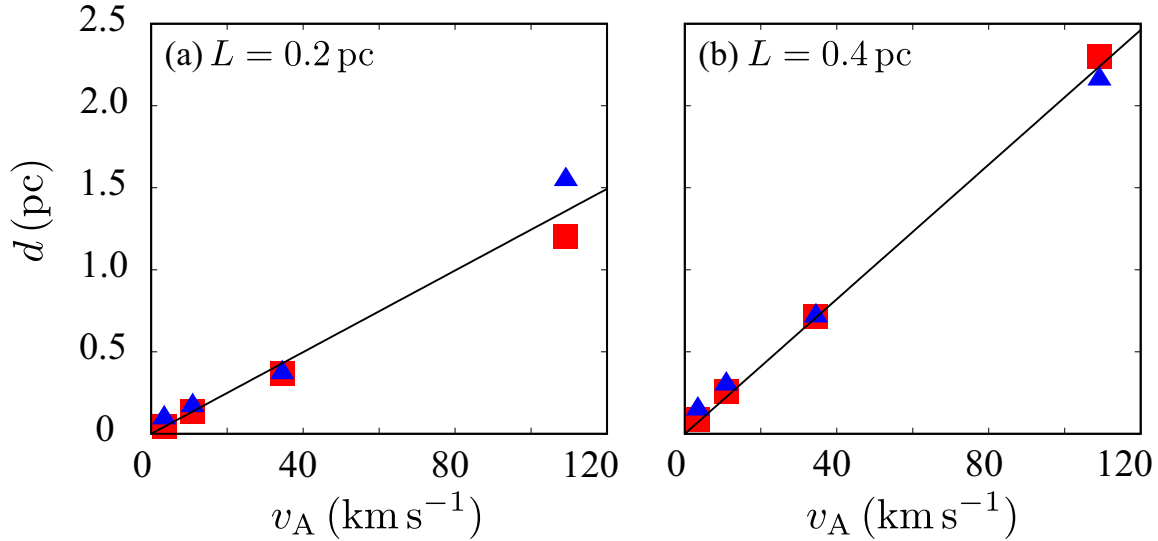}
\caption{Size of the accelerated gas
as a function of the Alfv\'en speed when $L=0.2\,{\rm pc}$ (a) and $L=0.4\,{\rm pc}$ (b).
The red squares represent the results for 
$n_0=10^5\,{\rm cm^{-3}}$, $10^4\,{\rm cm^{-3}}$, 
$10^3\,{\rm cm^{-3}}$, and $100\,{\rm cm^{-3}}$, 
while the magnetic field is kept constant at $B_{x0}=500\,\mu{\rm G}$. 
The blue triangles show the results for 
$B_{x0}=50\,\mu{\rm G}$, $158\,\mu{\rm G}$, $500\,\mu{\rm G}$, and $1.58\,{\rm mG}$, 
while the density is kept constant at $n_0=10^3\,{\rm cm^{-3}}$. 
The solid black lines shows the best-fitting lines
$d/{\rm pc}=0.012v_{\rm A}/{\rm km\,s^{-1}}$ (a) and $d/{\rm pc}=0.021v_{\rm A}/{\rm km\,s^{-1}}$ (b).
}
\label{fig4}
\end{figure*}

Figures \ref{fig4}(a) and \ref{fig4}(b) show the size of the ``Y'' shape 
on the $x$-$v_y$ map, $d$,
as a function of the Alfv\'en speed in the high-$B$ layer,
$v_{\rm A}=B_{x0}/\sqrt{4\pi\rho_0}$, when $L=0.2\,{\rm pc}$ and $0.4\,{\rm pc}$.
In order to survey the $v_{\rm A}$-dependence, 
we employ $n_0=10^5\,{\rm cm^{-3}}$, $10^4\,{\rm cm^{-3}}$, 
$10^3\,{\rm cm^{-3}}$, and $100\,{\rm cm^{-3}}$
while the magnetic field strength is set to $B_{x0}=500\,\mu{\rm G}$ (red squares).
In addition, we investigate the position--velocity structures for 
$B_{x0}=50\,\mu{\rm G}$, $158\,\mu{\rm G}$, $500\,\mu{\rm G}$, and $1.58\,{\rm mG}$
when the density is $n_0=10^3\,{\rm cm^{-3}}$ (blue triangles).
Here, the BH mass is kept constant $M_{\rm BH}=10^4 \,M_\odot$. 
The initial velocity $v_{y0}$ is adjusted to set the inflow velocity to $v_{\rm in}\sim 100\,{\rm km\, s^{-1}}$ before the high-$B$ layer reaches the BH.
Both Figures \ref{fig4}(a) and \ref{fig4}(b) show that 
the sizes are similar to each other if the Alfv\'en speeds have the same value, 
regardless of the combination of the density and the magnetic field strength.
The size increases in proportion to the Alfv\'en speed and the best-fitting lines are 
$d/{\rm pc}=0.012v_{\rm A}/{\rm km\,s^{-1}}$ and $d/{\rm pc}=0.021v_{\rm A}/{\rm km\,s^{-1}}$ for $L=0.2\,{\rm pc}$ and $0.4\,{\rm pc}$ (the solid black lines).



\begin{figure}[ht!]
\plotone{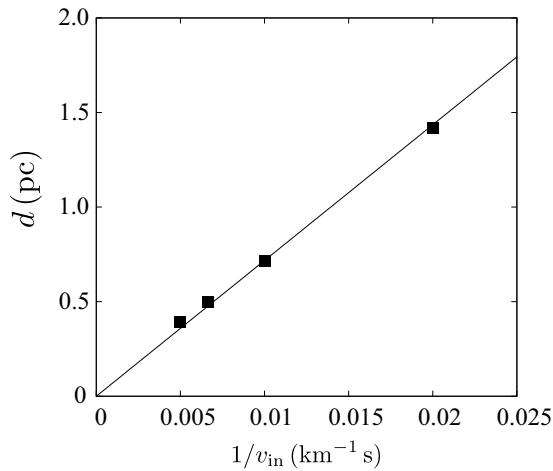}
\caption{Size of the accelerated gas
as a function of the inverse of the inflow velocity.
The black squares are the results for
$v_{\rm in}=50$, $100$, $150$, and $200\,{\rm km\,s^{-1}}$ 
(whose corresponding initial velocities are $v_{y0}=0$, $50$, $100$, and $150\,{\rm km\,s^{-1}}$).
The solid black line shows the best-fitting line of $d/{\rm pc}=72/(v_{\rm in}/{\rm km\,s^{-1}})$.
}
\label{fig5}
\end{figure}

Figure \ref{fig5} shows the size, $d$, 
as a function of the inverse of the inflow velocity, $1/v_{\rm in}$, when $L=0.4\,{\rm pc}$. 
We employ four different inflow velocities,
$v_{\rm in}=50$, $100$, $150$, and $200\,{\rm km\,s^{-1}}$, 
whose corresponding initial velocities are $v_{y0}=0$, $50$, $100$, and $150\,{\rm km\,s^{-1}}$
(black squares).
In these calculations, $M_{\rm BH}$, $n_0$, and $B_{x0}$ are 
the same as those of the fiducial model.
The size increases as the inflow velocity decreases
and this relation is fitted by 
$d/{\rm pc}=72/(v_{\rm in}/{\rm km\,s^{-1}})$ (the solid black line).
The decrease of the inflow velocity suppresses the ram pressure of the flow.
This leads the magnetic field lines drawing gentle curves and the large accelerated region.

\begin{figure}[ht!]
\plotone{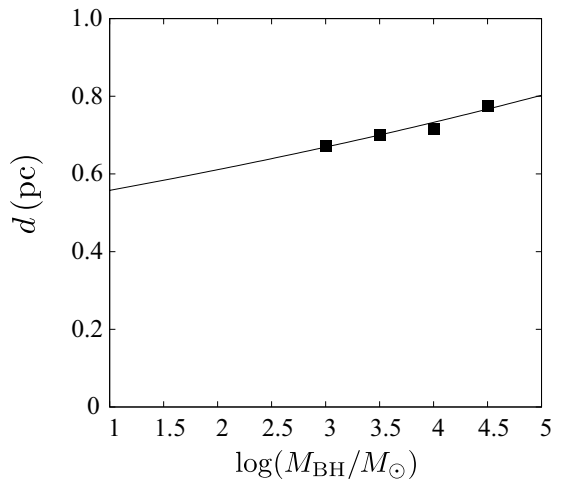}
\caption{Size of the accelerated gas
``Y'' shape on the $x$-$v_y$ map 
as a function of the BH mass.
The black squares show the results for
$M_{\rm BH}=10^3 \,M_\odot$, $10^{3.5} \,M_\odot$, $10^4 \,M_\odot$, and $10^{4.5} \,M_\odot$.
The best-fitting line of $d/{\rm pc}=0.51(M_{\rm BH}/M_\odot)^{0.040}$ is shown by the solid black line.
}
\label{fig6}
\end{figure}

Figure \ref{fig6} shows the $M_{\rm BH}$-dependence of the size, $d$, when $L=0.4\,{\rm pc}$. 
We calculate the position--velocity structures for 
$M_{\rm BH}=10^3 \,M_\odot$, $10^{3.5} \,M_\odot$, $10^4 \,M_\odot$, and $10^{4.5} \,M_\odot$ (black squares), 
while the other parameters are same as those of the fiducial model.
This figure shows that the size hardly depends on the BH mass 
at least in the range $10^3 \,M_\odot \leq M_{\rm BH}\leq 10^{4.5} \,M_\odot$.
The best-fitting line is $d/{\rm pc}=0.51(M_{\rm BH}/M_\odot)^{0.040}$ (the solid black line).
If we extrapolate this relation in the low-mass range, 
a ``Y''-shaped structure with a size of $\sim\! 0.5\,{\rm pc}$ 
is produced by a BH of $M_{\rm BH}\sim 10 \,M_\odot$. 

From Figures \ref{fig2}, \ref{fig4}(a), \ref{fig4}(b), \ref{fig5}, and \ref{fig6}, 
we find that the size, $d$, 
depends mainly on $v_{\rm A}$, $v_{\rm in}$, and $L$ 
besides having a weak dependence on $M_{\rm BH}$.
The size is approximately determined by 
$d=aL/{\mathcal M}_{\rm A}$,
where $a$ is a proportionality constant and ${\mathcal M}_{\rm A}$ is the Alfv\'en Mach number in the high-$B$ layer, 
${\mathcal M}_{\rm A}$=$v_{\rm in}/v_{\rm A}$.
We derive $a=5.1$ from the results of $M_{\rm BH}=10^4 \,M_\odot$.
The size of 
$L/{\mathcal M}_{\rm A}$
is the scale of the warped magnetic field lines 
when the high-$B$ layer passed $L$ after reaching the BH.
This is derived from the balance between the magnetic tension force and the ram pressure of the flow. 
The proportionality constant, $a(>1)$, shows 
an enhanced Alfv\'en speed and a decreased velocity around the BH.



\subsection{Comparison with Observations}
\label{subsec:o}
Our results show that the plunging of the BH into 
the high-$B$ layer
reproduces the characteristic ``Y'' shape on the position--velocity map 
of the Bullet in the W44 SNR.
Here, we quantitatively compare our results to the two objects, 
the W44 Bullet (Y17) and the small high-velocity compacts clouds (HVCCs) detected 
near the Galactic nucleus \citep[HCN--0.009--0.044 and HCN--0.085--0.094,][hereafter T17]{Takekawa17}
The small HVCCs have the velocity widths of $\Delta v \gtrsim 60\,{\rm km\,s^{-1}}$ 
and the sizes of $\sim\! 1\,{\rm pc}$. 
The high-velocity components originate from the dense molecular clouds. 
Although the ``Y'' shape is not resolved, 
the position--velocity structure of the small HVCCs is similar to that of the W44 Bullet.

\begin{figure}[ht!]
\plotone{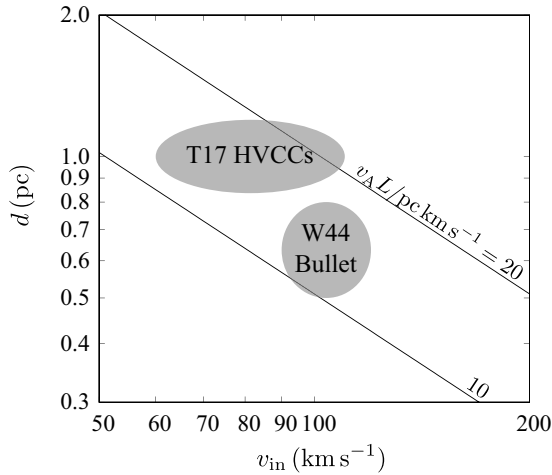}
\caption{
Comparison between our results and the observations on the $d$-$v_{\rm in}$ plane.
The solid lines show the constant $v_{\rm A}L$ lines based on the relation of $d=5.1L/{\mathcal M}_{\rm A}$.
The gray regions indicate the size and the velocity width of the Bullet (Y17) and the small HVCCs (T17). 
}
\label{fig7}
\end{figure}

In Figure \ref{fig7},
we compare the theoretical predictions based on our results and the observations on the $d$-$v_{\rm in}$ plane.
We plot the constant $v_{\rm A}L$ lines employing the relation 
$d=5.1L/{\mathcal M}_{\rm A}$
(the solid lines)
as well as the size and the velocity width of the Bullet and the small HVCCs (gray regions).
Here, on the basis of our simulations, 
we assume that the velocity width is comparable to the inflow velocity. 
We find that both the Bullet and the small HVCCs are located in the range 
$10\lesssim v_{\rm A}L\lesssim 20$, 
which can be rewritten as 
$5(L/{\rm pc})^{-1}(n/{\rm cm^{-3}})^{1/2}\lesssim B/\mu {\rm G} \lesssim 9(L/{\rm pc})^{-1}(n/{\rm cm^{-3}})^{1/2}$.

The value of $v_{\rm A}L$ can also be estimated from the observations. 
In the case of the Bullet, 
employing $B=500\,\mu{\rm G}$ \citep{Hoffman05}, $n=10^4\,{\rm cm^{-3}}$ (Y17), 
and $L=0.1\,{\rm pc}$ 
that is the thickness of thin filaments detected in W44 \citep{1993MNRAS.265..631J},
we obtain $v_{\rm A}L\sim 1.1\,{\rm pc\,km\,s^{-1}}$. 
This is consistent with the value predicted by our results within an order of magnitude. 
In the case of the small HVCCs, 
the typical magnetic field strength 
in the central region of our Galaxy 
is $B\sim 500\,\mu{\rm G}$
\citep[e.g.,][]{1996ARA&A..34..645M}
and the density of the molecular cloud is $n\sim 10^5\,{\rm cm^{-3}}$ \citep{1983A&A...125..136G}. 
We assume that $L$ is comparable to the HVCC size and is smaller than the size of the molecular cloud, 
$L\sim 1$--$5\,{\rm pc}$. 
As a result, we obtain $v_{\rm A}L\sim 3.5$--$17.3\,{\rm pc\,km\,s^{-1}}$. 
This is comparable to or slightly smaller than the theoretical prediction.

The difference between our results and the observations might be caused by 
the uncertainty of the coefficient, $a\sim 5.1$, in the relation
$d=aL/{\mathcal M}_{\rm A}$.
The coefficient, $a$, approximately corresponds to the ratio of the Alfv\'en speed 
at $r\sim R_{\rm BHL}$ to that of the high-$B$ layer at the initial condition. 
It is difficult to accurately evaluate the Alfv\'en speed 
in the vicinity of $r\sim R_{\rm BHL}$, 
because we ignore the dynamics in $r < R_{\rm BHL}$ and 
assume the simple conditions $n=0.1n_0$ and $v_{y}=0$. 
The magnetic field strength and the density near the BH would change 
if we consider the realistic magnetic structure around the BH 
and the feedback from the accretion flow such as radiative heating.
In that case, there is the possibility that the coefficient, $a$, increases by several times to ten times. 
and the size of the Bullet would be  well explained by the relation $d=aL/{\mathcal M}_{\rm A}$.


We found that the size of the accelerated region is almost independent of the BH mass (Figure \ref{fig6}).  
This result indicates that ``Y''-shaped position-velocity structure with $d\sim 0.5\,{\rm pc}$ can be reproduced by the plunging of a stellar mass BH. 
Here we estimate the total luminosity of the BH based on the BHL accretion model. 
Assuming $M_{\rm BH}=10\,M_\odot$, $n=10^4\,{\rm cm^{-3}}$, and $v_{\rm in}=100\,{\rm km\,s^{-1}}$, 
the total luminosity is estimated to be $L_{\rm X}=0.06 \dot M_{\rm BHL} c^2 \sim 2 \times 10^{34} \,{\rm erg\,s^{-1}}$, 
where $\dot M_{\rm BHL}$ is the BHL accretion rate \citep[e.g.,][]{2004NewAR..48..843E}.
This luminosity is consistent with the absence of the X-ray counterpart in {\it ROSAT} All Sky Survey \citep[RASS,][]{2009ApJS..184..138H}. 
If we assume the BHL accretion rate and standard accretion disk \citep{1973A&A....24..337S}, 
non-detection of the X-ray counterpart in the RASS indicates that the mass of the plunging BH is less than $\sim\! 100\,M_\odot$.  Note that the accretion rate would be smaller than $\dot M_{\rm BHL}$ because the magnetic field suppresses the mass accretion rate \citep{Lee14}.  In addition, the accretion disk might become radiation inefficient accretion flow \citep[RIAF,][]{1994ApJ...428L..13N}.  In such cases, a BH mass larger than $100\,M_\odot$ is acceptable. In order to calculate the accurate accretion rate and corresponding X-ray luminosity, the three-dimensional MHD and/or radiation hydrodynamics simulations in the small scale around the BH is necessary. Detection of an X-ray counterpart with modern X-ray imaging telescopes (such as {\it Chandra}) will provide a strong support for our scenario.  

As a first step of the theoretical approach to the Bullet, 
we performed the two-dimensional simulations on the $x$-$y$ plane ($z=0$ plane). 
In three-dimensional simulations,
the gas and magnetic field behaviors in the $z=0$ plane are expected to be similar to those in the two-dimensional simulations. 
This is because the gas and magnetic field lines far from the $z=0$ plane ($z \gg R_{\rm BHL}$) do not affect those in the $z=0$ plane. 
The three-dimensional MHD simulations of BHL accretion in a small computational domain show the bow shock similar to our results \citep{Lee14}, 
supporting that our two-dimensional simulations, at least qualitatively, well reproduce the gas and magnetic field behaviors in the $z=0$ plane. 
In the three-dimensional simulations, the magnetic pressure in the $z$-direction might lower the density and magnetic field strength near the BH, 
and thereby reduce the size of the accelerated region in $z=0$ plane. 
Investigating these three-dimensional effects must be important and interesting future work. 

In our simulations, we assumed a simple parallel-magnetic-field layer, 
although magnetic fields in molecular clouds are not highly ordered. 
The unordered fields may produce the asymmetric ``Y'' shape and the size of the accelerated region might slightly change. 
More accurate measurements of the magnetic field configuration near the Bullet and the simulations employing realistic settings may be 
interesting future works.

\section{Conclusions}\label{sec:conclusions}
Performing the MHD simulations, 
we investigated the gas dynamics around 
a BH plunging into a molecular cloud with 
a parallel-magnetic-field.
We found the following results:
\begin{enumerate}
\item The MHD effects enlarge the accelerated region compared to the native shooting model, and the acceleration region expands within $|x|\lesssim 0.36\,{\rm pc}$ when $L=0.4\,{\rm pc}$ for the fiducial model ($M_{\rm BH}=10^4 \,M_\odot$, $B_{x0}=500\,\mu {\rm G}$, $n_0=10^3\,{\rm cm^{-1}}$, and $v_{\rm in}=100\,{\rm km\,s^{-1}}$).
\item When $L=0.4\,{\rm pc}$, the accelerated gas exhibits a ``Y'' shape with a size of $d\sim 0.72\,{\rm pc}$ and a velocity width of $\Delta v \sim 115\,{\rm km\,s^{-1}}$ on the $x$-$v_y$ map 
for the fiducial model.
\item The size of the accelerated gas increases 
in proportion to the time (the distance traveled in the layer, $L$).
\item Our simulations show that the size of the ``Y'' staple is almost independent of the BH mass 
and stellar mass BH ($M_{\rm BH}\lesssim 100\,M_\odot$) is preferred for the Bullet if we assume the BHL accretion rate and the standard accretion disk.
\item The size of the ``Y'' shape is approximately determined by $d=5.1L/{\mathcal M}_{\rm A}$. 
\item Our model can reproduce the ``Y'' shape and 
the velocity width on the position--velocity map of the W44 Bullet. 
\item The size of the ``Y'' shape expected from the model is consistent with that of the Bullet within one order of magnitude. 
\item Our model can explain the velocity width and the size of the small HVCCs in the Galactic center.
\end{enumerate}
 
The foregoing results support the shooting model of the Bullet and the small HVCCs 
and indicate that the MHD effects are necessary to reproduce the size of the accelerated gas.
There is a possibility that the plunging of the isolated BH into the molecular cloud be 
responsible for the formation of the extraordinary high velocity component.

\acknowledgments

Numerical computations were carried out on Cray XC30 at the Center for Computational Astrophysics, 
National Astronomical Observatory of Japan. 
This work is supported in part by JSPS Grant-in-Aid for Scientific Research (B) (15H03643 T.O., 15K05036 K.O), for Young Scientists (17K14260 H.R.T), and for Research Fellow (15J04405 S.T). 
This research was also supported by MEXT as ``Priority Issue on Post-K computer'' (Elucidation of the Fundamental Laws and Evolution of the Universe) and JICFuS.

\end{document}